\documentstyle[11pt]{article}
\newcommand{\alAu}{$\alpha$~$+$~Au~}
\newcommand{\alfAu}{$\alpha$(5 GeV/u)~$+$~Au~}
\newcommand{\Elab}{$E_{\rm lab}$}

\newcommand{\vr}{\mbox{\boldmath $r$}}

\title{\bf
Multifragmentation in\\
{\boldmath ${\alpha}$}(5GeV/u) + Au Reactions}

\author{
Tomoyuki Maruyama\\
Advanced Science Research Center,\\
Japan Atomic Energy Research Institute,\\
Tokai, Ibaraki 319-11, Japan }
\date{}

\begin{document}

\maketitle

\begin{abstract}

We simulate the fragmentation processes in the \alAu collisions at
the bombarding energy 5 GeV/u using the simplified RQMD approach.
The statistical decay calculation is connected to obtain the final state.
 From the simulation we find that the residual nucleus created by this
collision has a annular eclipse shape, and then expands more strongly to
the transverse direction than to the beam-direction.
Then the angular-distribution of the intermediate mass fragments
has a sideward peak; this result explains the experimental result.

\end{abstract}

\bigskip
\vspace{5cm}

\begin{center}
Talk in the 2nd Joint RIKEN(Japan)/INFN(Italy) Simposium
\end{center}

\newpage

The main aim of the high and intermediate energy light- and heavy- ion
physics is to investigate properties of nuclear matter under extreme
conditions, especially to determine the nuclear equation of state (EOS).
Any answer about properties of hot and dense matter must rely on
the comparison of the experimental data with theoretical predictions.
Among these, the numerical simulation approaches, such as BUU \cite{Bert,BUU1}
and the Quantum Molecular Dynamics (QMD) \cite{Aich}, are most promising
in describing the time-evolution of the complex system.
There are two important ingredients of transport theories:
a mean-field for nucleons and in-medium baryon-baryon
cross-sections that account for elastic and inelastic channels.
By searching the mean-field which can fit the experimental data,
one expects to know the nuclear matter properties \cite{BUU3}.

Multifragmentation attracts attention as one of most important aspects
of light- and heavy- ion reactions in the high energy region \cite{expfr}.
It is speculated that the decay of a highly excited nuclear system
carries the information about the nuclear EOS and
the liquid-gas phase transition of low density nuclear matter.

Recently the KEK experimental group has reported very interesting
results for the distribution of the intermediate mass fragments (IMF)
in proton and alpha induced reactions at 12 GeV and at 5 GeV/u,
respectively \cite{KEK}.
The experimental data for the IMF angular-distributions have clear
peaks at $\theta_{lab} = 70^\circ$.

In this reaction we can easily consider that some high energy pions,
protons and light fragments such as deutron are emitted forwards
immediately after starting the collision, and then that
IMF are formed through the thermal decay of a hot residual nucleus.
If the hot residual nucleus decays isotropically, the angular-distribution
should have a forward peak.
In the actual process, hence, the residual nucleus is expected to have
a exotic shape and to expand more strongly to the transverse direction than
to the beam-direction.

Now the QMD approach is the most useful one for the theoretical study of
fragmentation.
T. Maruyama et al. \cite{Toshi} have succeeded to reproduce experimental
data of heavy-ion collisions around several ten MeV/u by using QMD with the
statistical decay code \cite{SDecay}.
Since the reaction process cannot be experimentally determined without
a theoretical analysis,
it is natural to apply this QMD approach to the high energy region.

For the purpose of studying the fragmentation process, we need to describe
the time evolution of the phase-space distribution until its expansion stage
in the simulation.
At relativistic energy all nuclei and fragments must, in principle, hold
the consistent phase-space distribution under the Lorentz transformation.
Otherwise we mainly have two kinds of troubles in the actual calculation
as follows.
First the increase of the initial density due to the Lorentz contraction
strengthen repulsion through the density-dependent force and causes the
unphysical instability of initial nuclei \cite{Sorge}.
Second we cannot correctly evaluate the internal energies of fast-moving
fragments at the end of the usual QMD calculation because this approach uses
the relativistic kinematics and the non-relativistic mean-field
which is variant under the Lorentz transformation \cite{TOMO2}.
In fact, these effects largely influence the multiplicity of alpha even
at \Elab $\sim$ 1 GeV/u \cite{TOMO2}.
Hence a fully Lorentz covariant transport approach is desired
in the relativistic energy region.

The Relativistic QMD (RQMD) approach \cite{Sorge,MARU1} must be the most
useful theoretical model for this purpose; it is formulated to describe
the interacting $N$-body system in a fully Lorentz covariant way based on
the Poincare-Invariant Constrained Hamiltonian Dynamics \cite{PICHD}.
The position $q_i$ and momentum coordinate $p_i$ of the $i$-th nucleon
are defined as four-dimensional dynamical variables and the functions of
the time evolution parameter $\tau$.
The on-mass-shell constraints are given by
\begin{equation}
H_i \equiv p_i^2 - m_i^2 - 2 m_i {\tilde V_i} = 0 ,
\end{equation}
where $m_i$ and ${\tilde V_i(q_j,p_j)}$ is a mass and a Lorentz-scalar
quasi-potential.
The detailed form of the quasi-potential is determined by
the requirement that it is corresponding to the non-relativistic
mean-field in the low energy limit \cite{Sorge,MARU1}.
Whereas in the non-relativistic framework the argument of the potential
is a square of the relative distance between two nucleons
$\vr^2_{ij}$, in the RQMD we take it as a square of the relative
distance at the rest frame of their CM system,
\begin{equation}
- q_{{\rm T}ij}^2 =
- q_{ij}^2 + \frac{ ( q_{ij} \cdot p_{ij} )^2 }{ p_{ij}^2 }.
\end{equation}
with
\begin{eqnarray}
q_{ij} & = & q_i - q_j ,
\nonumber \\
p_{ij} & = & p_i + p_j .
\end{eqnarray}
Through this change from $\vr_{ij}^2$ to $- q_{{\rm T}ij}^2$, there appears
the total momentum-dependence of two nucleon force while the
usual momentum-dependence is only on the relative momentum.
This causes the direction-dependent forces and the Lorentz contracted
phase-space distribution of fast moving matter.

In this formulation the time coordinate $q_i^0$ is distinguished from
the time evolution parameter $\tau$.
In Ref. \cite{Sorge,MARU1} time-ficxation has been given to require equal time
coordinates of two colliding particles in their center-of-mass system.
Using the above on-mass-shell and the time-fixation constraints,
we can describe the Lorentz covariant motions of nucleons.

These constraints are chosen to be completely consistent
with the non-relativistic framework in the non-relativistic limit
($m_i \rightarrow \infty$).
Hence this RQMD approach must be very useful to obtain the fragment
distribution theoretically because the excitation energy of fast moving
nuclear fragments can be evaluated without any ambiguity caused by
the boost \cite{TOMO2}.

However there are two kinds of  troubles to apply the RQMD approach to
our analysis of this \alAu collision.
One is that this code spends too long CPU time.
The other is that it is not easy to satisfy the energy conservation
after the meson production and absorption.
Changing the number of particles largely breaks the time fixation.
Resolving the constraints, the on-mass shell condition and the time-fixation,
makes the discontinuity of particle coordinates, and changes the total energy.
Especially the latter problem is too serious to study the fragmentation
because IMF multiplicities are almost determined by the excitation energy
of the residual nucleus

In order to avoid these difficulties we simplify the RQMD approach by
taking the time fixations to equalize all time coordinates of baryons and
mesons.
Though this approximation breaks the Lorentz covariance
in two body collisions in principle,
this breaking is not so serious in the several GeV/u energy region
\cite{GyWolf}.
Of course the meson production and absorption do not change the time
coordinate and do not break the energy conservation.
For this reason we will simulate the dynamical stage of
\alfAu collisions with this simplified RQMD (RQMD/S) and
investigate the origin of the experimental results for
the IMF angular-distribution.

The actual calculations are made in the following ways.
First the initial distribution at rest is made by the cooling
method \cite{Cooling} and boosted according to the bombarding energy.
Second we perform the RQMD/S calculations and obtain dynamical fragment
distribution.
Third we boost each dynamical fragment to its rest frame and evaluate
its excitation energy.
Finally we calculate the statistical decay \cite{SDecay} from
the dynamical fragments and obtain the final fragments distribution.

As a mean-field we use a Skyrme-type interaction with HARD EOS (the
incompressibility $K$ = 380 MeV) which is parameterized in Refs.
\cite{Bert,Aich},
where the density-dependent part is treated in the way of Ref. \cite{Cooling}.
In addition the symmetry force and the Coulomb force is also introduced to get
a correct isospin of a fragment in the simulation.
As a cross-section of two baryon collisions we use Cugnon's
parameterization \cite{Cug} for a elastic channel and Wolf's  parametrization
\cite{GyWolf,Giessen} for inelastic channels including three baryonic
resonances: $\Delta, N^{*}(1440)$ and $N^{*}(1535)$.
These resonances decay into nucleons and mesons ($\pi$ and $\eta$)
\cite{Giessen}.

We show  the baryon and pion distributions in the orbital space every
4 fm/c time step, projecting on the $xz$- plane (restricted with positions
$|y| < 1$ fm) in Fig. \ref{tvxz}, and projecting on the $xz$- plane
(restricted with positions $|z| < 1$ fm) in Fig. \ref{tvxy})
There twenty simulations have been performed for the impact-parameter
$b = 0$ fm,
and full circles and crosses indicate baryons and mesons, respectively

Around the time step $t = 12$ fm/c a lot of pions are produced and propagate
forwards.
The empty region in the center appears at $t = 20$ fm/c;
namely the residual nucleus with the annular eclipse shape is constructed
through this reaction.
After that the residual nucleus slowly expands sidewards, and
finally nucleons gather, connect and form fragments.
Apparently this fragmentation process is the multifragmentation.

In Fig. \ref{angd}, next, we show the angular-distribution of pions and
four kinds of fragments with charge $Z = 1$ (full triangles),
$Z = 2$ (open squares), $Z =3,4$ (open diamonds) and
$5 \leq Z \leq 20$ (full circles),
where  events are restricted with the impact-parameter $b < 6$ fm.
Pions and fragments with $Z = 1$ have a clear forward peak, while this
forward peaking becomes mild as fragments become heavier.
Then the angular-distribution of  fragments with $5 < Z < 20$  has a clear
peak at $\theta_{lab} = 80^\circ$, and this result agrees with experimental
one qualitativerly.

Of course all collisions do not construct the annular eclipse nucleus;
the shape of the residual nuclei depend	 on the impact-parameter.
The annular eclipse shape appears only in the cases of the central and
semi-central collisions.
The peripheral collisions does not cause the multifragmentation and give the
forward-peaking angular-distribution.
They do not contribute to the cross-section largely, however, because the IMFs
cannot be often produced.
In the middle impact-parameter region, on the other hand, the residual nucleus
has a partial eclipse and expands sidewards.
Such a process also makes the side-directed peak of the IMF
angular-distribution.
Namely the observed cross-section are given from both the annular eclipse and
partial eclipse processes.

Furthermore we should give one more comment.
In the residual nucleus with the annular eclipse shape the mean-field is
attractive and work to disturb the side-directed expansion.
If the side-directed force given by nucleons and mesons passing through
the target is weaker than the attractive mean-field, the large
compound nucleus without the central empty region is
produced and  IMF is created through the evaporation and the binary fission.
Since the residual nucleus moves almost on the beam-direction, the IMF
angular-distribution must have the forward peak in the above process.

 From these results we can conclude that the $\alpha$ + Au reaction at
5 GeV/u constructs a hot nuclear system with the annular eclipse shape in the
central collision and the partial eclipse shape in the non-central collisions.
Then this residual nucleus expands sidewards and causes a multifragmentation.
As a result of this process the IMF angular-distribution has a side-directed
peak.

If we choose events in the central collisions,
we will be able to get the detailed information of hot nuclei with the annular
eclipse shape.

%\newpage

\newpage

\noindent
{\large\bf Figure captions}\hfill
\vspace{1em}

\noindent
{\bf Fig.~1:}
The distribution of baryons and pions in the configuration space ($xz$- plnae),
restricted $|y| < 1$ fm, at time steps at 4, 8, $\cdots$ 36 fm/c
in a \alfAu collisions for the impact-parameter $b = 0$ fm.
Full circles and crosses indicate baryons and mesons, respectively.

\vspace{1em}\noindent
{\bf Fig.~2:}
The distribution of baryons and pions in the configuration space ($xy$- plnae),
restricted $|z| < 1$ fm, at time steps at 4, 8, $\cdots$ 36 fm/c
in a \alfAu collisions for the impact-parameter $b = 0$ fm.

\vspace{1em}\noindent
{\bf Fig.~3:}
The angular-distributions of pion (crosses) and fragments:
$Z = 1$ (full triangles), $Z = 2$ (open squares), $Z = 3,4$ (open diamonds)
and $5 \leq Z \leq 20$ (full circles).
The cross-section for the last fragment is multiplied by 10.
Events are restricted with  the impact-parameter $b < 6$ fm.

\eject
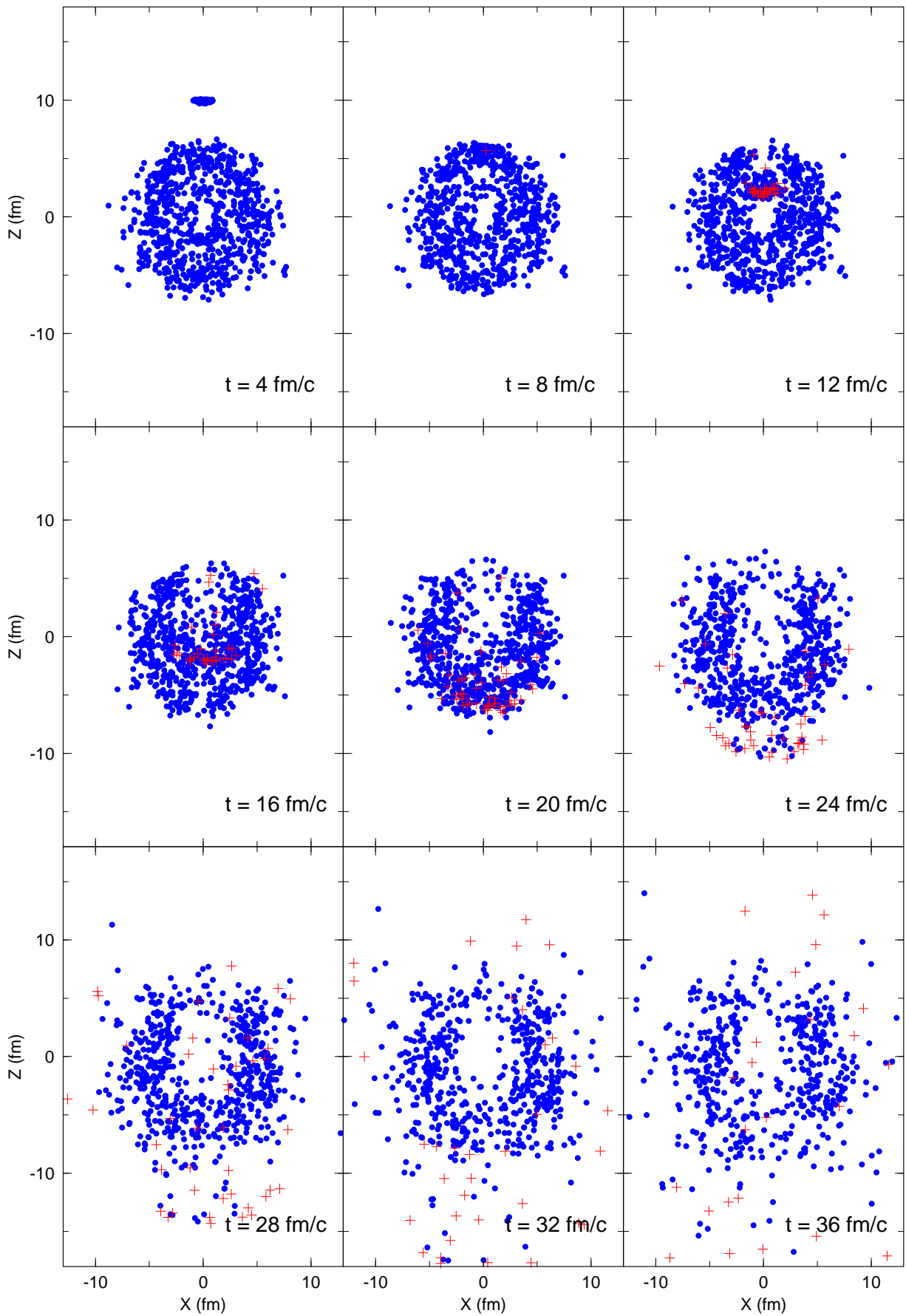
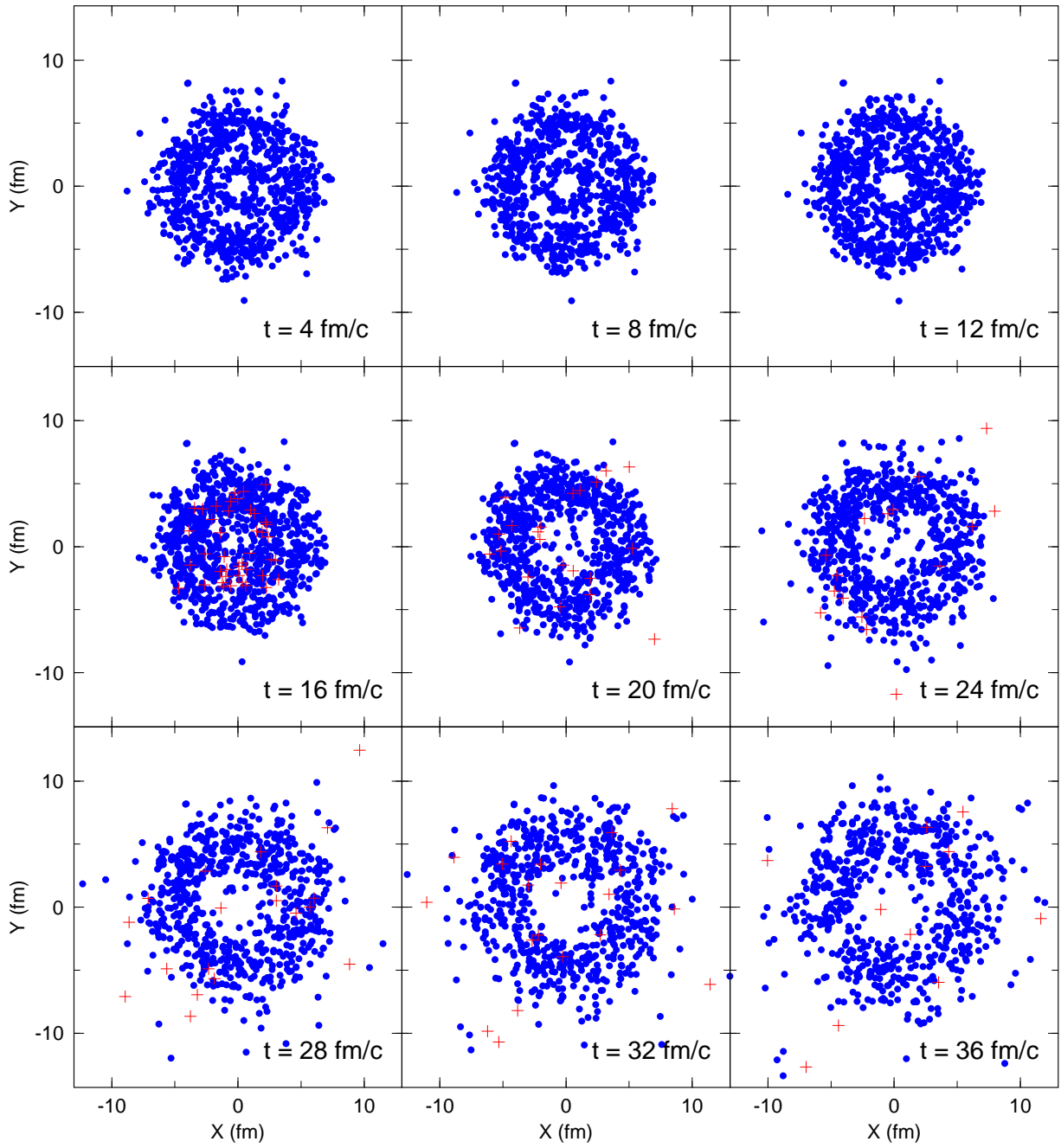
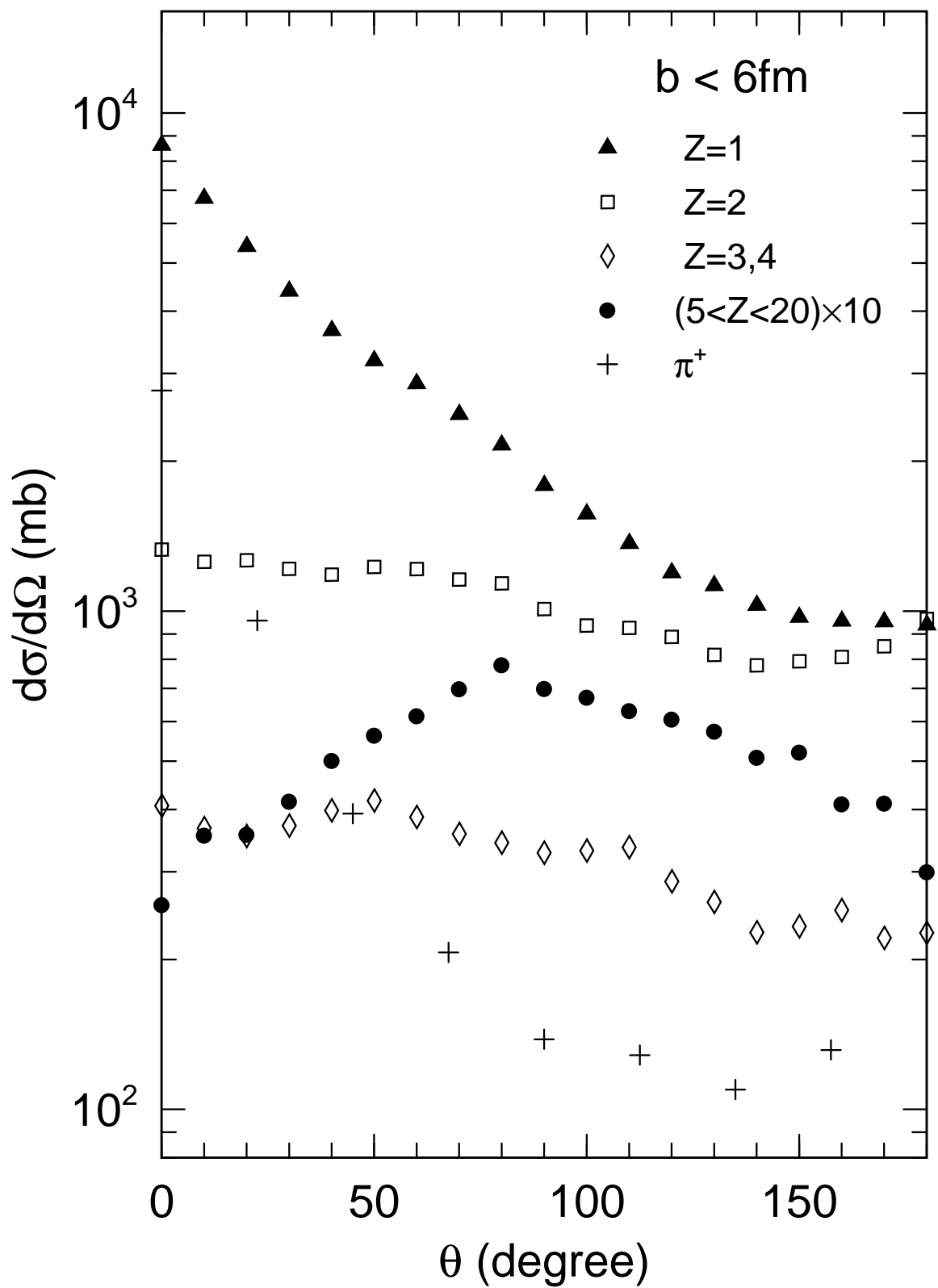
\end{document}